# On Return Rate Implied by Behavioural Present Value

Krzysztof PIASECKI,

Department of Operations Research, Poznań University of Economics, Poland,

e-mail: k.piasecki@ue.poznan.pl

The future value of a security is described as a random variable. Distribution of this random variable is the formal image of risk uncertainty. On the other side, any present value is defined as a value equivalent to the given future value. This equivalence relationship is a subjective. Thus follows, that present value is described as a fuzzy number, which is depend on the investor's susceptibility to behavioural factors. All above reasons imply, that return rate is given as a fuzzy probabilistic set. The basic properties of such image of return rate are studied. At the last the set of effective securities is distinguished as a fuzzy set.

## Introduction

Typically, the analysis of properties of any security is kept, as analysis of return rate properties. Any return rate is increasing function of future value and decreasing function of present value. The future value of a security is presented as a random variable. Distribution of this random variable is formal image of uncertainty risk. On the other side, any present value is defined as such current value which is equivalent to the given future value. This equivalence relation is subjective by nature, because it depends on the investor's susceptibility to internal and external behavioural factors. It implies that the present value can be deviated from its observed market price out of influence of behavioural factors. States of the behavioural environment are defined imprecisely. For this reason a present value may be given as a fuzzy number. An example of such present value is described and discussed by Piasecki (2011). Then the return rate of is given as a probabilistic fuzzy number. Basic properties of such return rate will be investigated in this paper. The main goal of our considerations will be to define a set of effective securities.

### 1. Imprecise assessment of return rate

Let us assume that time horizon $t > 0$ of investment is fixed. Then considered security is determined by two values:

- anticipated future value $V_t \in \mathbb{R}^+$,
- assessed present value $V_0 \in \mathbb{R}^+$.

The basic characteristics of benefits by ownership this instrument is a return rate $r_t$ given by the identity

$$r_t = r(V_0, V_t). \tag{1}$$

In the general case, the function: $r: \mathbb{R}^+ \times \mathbb{R}^+ \to \mathbb{R}$ is a decreasing function of the present value and a increasing function of future value. It implies that for any present value $V_0$ and future value $V_t$ we can determine reciprocal function $r_0^{-1}(\cdot, V_t): \mathbb{R} \to \mathbb{R}^+$. In the special case we have here:

- simple return rate

$$r_t = \frac{V_t - V_0}{V_0} = \frac{V_t}{V_0} - 1. \tag{2}$$

- logarithmic return rate

$$r_t = ln \frac{V_t}{V_0}. \tag{3}$$

The future value of investments $V_t$ is at risk of uncertainty about the future state of affairs. Formal model of this uncertainty is presentation future value as a random variable $\tilde{V}_t: \Omega = \{\omega\} \to \mathbb{R}^+$. The set $\Omega$ is a set of elementary states of the financial market. In the classical approach to the problem of return rate determination, present value of a financial instrument is identified with the observed market price $\check{C}$. Then return rate is a random variable which is at uncertainty risk of uncertainty. This random variable is determined by the identity

$$\tilde{r}_t(\omega) = r\left(\check{C}, \tilde{V}_t(\omega)\right). \tag{4}$$

In practice of financial markets analysis, the uncertainty risk is usually described by probability distribution of returns rate. At the moment we have an extensive compendium of knowledge on this subject. Let us assume that this probability distribution is given by cumulative distribution function $F_r: \mathbb{R} \to [0; 1]$. Then probability distribution of future value is described by cumulative distribution function $F_V: \mathbb{R}^+ \to [0; 1]$ given as follows

$$F_V(x) = F_r\left(r_t(\check{C}, x)\right). \tag{5}$$

Cumulative distribution function $F_V: \mathbb{R} \to [0; 1]$ describes the probability distribution of future value. Assessment of this variable is based on objective measurement only. It means that the cumulative distribution function of future value is independent of the way of determining the present value.

Piasecki (2011) shown that the present value may be at imprecision risk. Mentioned imprecision risk was determined by behavioural premises. Imprecise assessed initial value is represented by its membership function $\mu: \mathbb{R}^+ \to [0; 1]$. Then the return rate is at risk of coincidence uncertainty rand imprecision. According to the Zadeh extension principle, for each fixed elementary state $\omega \epsilon \Omega$ of financial market, membership function $\rho(\cdot, \omega): \mathbb{R} \to [0; 1]$ of return rate is determined by the identity

$$\rho(r, \omega) = \max\{\mu(y): y \epsilon \mathbb{R}^+, r = r(y, \widetilde{V}_t(\omega))\} = \mu\left(r_0^{-1}(r, \widetilde{V}_t(\omega))\right). \tag{6}$$

It means that considered return rate is represented by fuzzy probabilistic set. The notion of probabilistic fuzzy set was suggested and studied by Hiroto (1981). For this reason, these sets are also called Hiroto's sets. In special cases we have here:

- for the simple return rate

$$\rho(r, \omega) = \mu\left((1 + r)^{-1} \cdot \widetilde{V}_t(\omega)\right) \tag{7}$$

- for the logarithmic return rate

$$\rho(r, \omega) = \mu\left(e^{-r} \cdot \widetilde{V}_t(\omega)\right) \tag{8}$$

For above described return rate we determine the parameters of its distribution. We have here:

- distribution of expected return rate

$$\rho(r) = \int_{-\infty}^{+\infty} \mu\left(r_0^{-1}(r, \widetilde{V}_t(\omega))\right) dF_V(y); \tag{9}$$

- expected return rate

$$\bar{r} = \frac{\int_{-\infty}^{+\infty} \int_{-\infty}^{+\infty} r \cdot \mu\left(r_0^{-1}(r, \widetilde{V}_t(\omega))\right) dF_V(y) dr}{\int_{-\infty}^{+\infty} \int_{-\infty}^{+\infty} \mu\left(r_0^{-1}(r, \widetilde{V}_t(\omega))\right) dF_V(y) dr}. \tag{10}$$

Distribution of expected return rate $\rho \epsilon [0,1]^{\mathbb{R}}$ is a membership function of fuzzy subset $\widetilde{R}$ in the real line. This subset $\widetilde{R}$ is called fuzzy expected return rate. This rate represents both rational and behavioural aspects in the approach to estimate the expected benefits. As the assessment of the risk uncertainty we take the variance of return rate

$$\sigma^2 = \left(\int_0^{+\infty} \int_{-\infty}^{+\infty} v(r, y) dF_V(y) dr\right)^{-1} \cdot \int_0^{+\infty} \int_{-\infty}^{+\infty} r \cdot v(r, y) dF_V(y) dx \tag{11}$$

where

$$\nu\left(x, \tilde{V}_t(\omega)\right) = \begin{cases} max\{\rho(\bar{r} + \sqrt{x}, \omega), \rho(\bar{r} - \sqrt{x}, \omega)\} & x \geq 0 \\ 0 & x < 0 \end{cases} \quad . \quad (12)$$

Detailed analysis of these relationships show that determined in this way variance describes both rational and behavioural aspects of safety assessment of invested capital.

Similarly as in the case of precisely defined return rate, there are such probability distributions of future value for which return rate variance does not exist. Then we replace this distribution with distribution truncated on both sides, for which the variance always exists. This procedure finds its justification in the theory of perspective introduced by Kahneman and Tversky (1979). Among other things, in this theory is described the behavioural phenomenon of extremes rejection.

Despite this modernization, in the proposed model can be used without changes all rich empirical knowledge about probability distributions of return rate. This fact expands the possibility of real applications. It is a highly advantageous feature of the proposed model.

### 3. The three-dimensional image of the risk

In the classical portfolio theory given by Markowitz (1952) the normative investment strategy is maximization of expected return rate $\bar{r}$ while its variance σ is minimizing. In this situation, each security is represented by pair $(\bar{r}, \sigma^2)$. This pair represents reasonable circumstances for securities evaluation. There are implicitly assumed that the returns have normal distributions.

In this chapter the basic image $(\bar{r}, \sigma^2)$ of the financial instrument is replaced by a pair $(\tilde{R}, \sigma^2)$ which also taking into account behavioural aspects of decision making in finance. In this way, we increase the cognitive value of description of the security. However, this increase in utility of this description has a price. This price is a disclosure of imprecision risk of return rate. Imprecision risk is composed of the ambiguity risk and indistinctness risk.

The ambiguity risk is due the lack of clear recommendation one alternative between the various given alternatives. In accordance with suggestion given by Czogala, Gottwald and Pedrycz (1982), we will evaluate the ambiguity risk by energy measure $d(\tilde{R})$ of fuzzy expected return rate $\tilde{R}$. This measure is determined by the identity

$$\delta = d(\tilde{R}) = \frac{\int_{-\infty}^{+\infty} \rho(x)dx}{1+\int_{-\infty}^{+\infty} \rho(x)dx} \cdot \quad . \quad (13)$$

Indistinctness risk is due to the lack of explicit distinguishing amongst the given information and its negation. According to suggestion given by Gottwald, Czogala and Pedrycz, (1982), we will evaluate the indistinctness risk by entropy measure $e(\tilde{R})$ of fuzzy expected return rate $\tilde{R}$. This measure is described as follows

$$\varepsilon = e(\tilde{R}) = \frac{\int_{-\infty}^{+\infty} min\{\rho(x), 1-\rho(x)\}dx}{1+\int_{-\infty}^{+\infty} min\{\rho(x), 1-\rho(x)\}dx}. \tag{14}$$

In this situation, for each fuzzy expected return rate $\tilde{R}$ we assign three-dimensional vector $(\sigma^2, \delta, \varepsilon)$. This vector is a image of the risks which is understood, as the composition of the risks of uncertainty, ambiguity and indistinctness.

The uncertainty risk follows from the lack of investor knowledge of future states of the financial market. Lack of this knowledge implies that any investor is not sure of future profits or losses. Properties of this risk are discussed in the rich literature.

We put here the question whether is relevant the imprecision risk to the analysis of investment processes. Investor takes some responsibility for making decision making on advisers or applied prognostic tool. For this reason, investor restricts its choice of investment decisions for alternatives recommended by consultants. In this way the investor minimizes his individual responsibility for financial decision making. This problem was widely discussed by Piasecki (1988).

Increase of ambiguity risk means that the number of recommended investment alternatives investment increases too. This increases the chance of selecting the recommended alternative, which is encumbered with loss of lost opportunities.

Increase in imprecise risk means that differences between delimitation between recommended and unrecommended alternatives are more blurred. It implies increase in chance of choice unrecommended alternatives.

These observations show that the increase in imprecision risk makes investment conditions noticeably worse. So, imprecision can be considered as a risk which is relevant to investment process.

Using three-dimensional image of risk $(\sigma^2, \delta, \varepsilon)$ makes easier management of imprecision risk. Here it is desirable to minimize each of the three risk assessments. Using three-dimensional image of risk enables investigation of relationships between different types of risk. Here we can to observe empirical interaction between risks. Moreover, there is a formal correlation between uncertainty risk and the ambiguity risk. The number of recommended alternatives increases with ambiguity risk. In this way, is becoming more certain that between the recommended alternatives is the best investment decision. This means that the uncertainty risk decreases. In summary, the uncertainty risk and the ambiguity risk are negatively correlated.

In comparison with the classical Markowitz theory imprecision is a new aspect of risk assessment. We put here the question whether is appropriate such extension of risk assessment. The usefulness of taking into account imprecision in risk study is well justified by following three arguments.

At the first, always is possible to reduce the uncertainty risk of forecast may be reduced by appropriate manipulation of lowering the forecast precision.

At the second, if we take into account imprecision risk then we will reject investment alternatives which are attractive from the viewpoint of classical Markowitz theory, but unfortunately information gathered about them are highly imprecise.

At the third, from the viewpoint of classical Markowitz theory and its implications, in financial markets practice we meet with many anomalies.

Seeing these paradoxes is the starting point for the development of behavioural finance. In this paper we show how the consideration of imprecision risk leads to normative theory explaining financial market paradoxes.

## 4. Financial effectiveness

If for given variance the security has a maximum expected return rate, then it is called an effective one. In the classical portfolio theory Markowitz assumed that the distribution of return rates is gaussian. Then the set of effective securities is given as the upper branch of the Markowitz curve which is called effective securities curve.

The set of effective securities can be specified also by means of theory of multicriteria comparison. Using this approach we can dispense with the assumption that the probability distribution of return rates is gaussian. Using this approach we define two preorders on the set of all securities. These preorders are the maximization expected return rates and the minimization variance. The set of effective securities is described as the Pareto optimum set for multicriteria comparison defined by above preorders. If in addition we assume here that the return rates distribution is gaussian, then the set of effective securities coincides with the upper branch of Markowitz curve. This means that the set of effective securities is a generalization of the concept of effective securities curve defined on the basis of the classical Markowitz theory.

Any investment in effective security is the investment in security guarantying maximum returns with minimal risk of capital loss. This is a standard investor's goal in normative theories of financial market. This causes some difficulties in applications, because of investors typically invest in securities lying outside the set of effective one. So, from the viewpoint of these theories, they invest in inefficient securities. At the same time, these investors declare investing in efficient securities as its normative goal. In this way we find a paradox of real financial market.

Mentioned above paradox occurs very often. This fact cannot be explained by lack of sufficient knowledge of the real processes occurring in the financial markets and economic environment. Increasing professionalization of investor activity and fast development of informatics imply that full access to market information and its processing capacity is available to all investors who manage the vast majority of exchange trading volume.

Considered paradox may be explained in the following way. The normative aim of investing in effective securities is declared by investors who invest only in securities similar to the effective one. Degree of effectiveness of given security is equal to the degree of its similarity to an effective one. In practice this means that almost every commercially available security is effective to some extent. On the other hand, inefficient financial instrument ceases to be object of stock-exchange turnover. All these explain the paradox of divergence between the normative investor's purpose and the real goal of investment strategy. Investors always act in manner similar to effective one.

Let us consider the normative model of investors' activity. The set of all securities is denoted by the symbol $\mathbb{Y}$. The security $\breve{Y} \epsilon \mathbb{Y}$ is represented by the pair $\left(\tilde{R}_Y, (\sigma_Y^2, \delta_Y, \varepsilon_Y)\right)$, where the individual symbols mean:

- $\tilde{R}_Y$ is fuzzy expected rate of return on security $\breve{Y}$,
- $\sigma_Y^2$ is the variance rate of return on security $\breve{Y}$,
- $\delta_Y$ is the energy measure of fuzzy expected return rate $\tilde{R}_Y$,
- $\varepsilon_Y$ is the energy measure of fuzzy expected return rate $\tilde{R}_Y$.

The fuzzy expected return rate $\tilde{R}_Y$ is defined by distribution of expected return rate $\rho_Y \epsilon [0,1]^{\mathbb{R}}$. On the set of fuzzy real numbers $\mathcal{F}(\mathbb{R})$ define the relation $\tilde{K} \succcurlyeq \tilde{L}$, which reads:

*Fuzzy real number $\tilde{K}$ is greater or equal to fuzzy real number $\tilde{L}$.*

This relation is a fuzzy preorder defined by such membership function $\nu_Q: \mathcal{F}(\mathbb{R}) \times \mathcal{F}(\mathbb{R}) \to [0,1]$ which fulfils the condition

$$\nu_Q(\tilde{R}_Y, \tilde{R}_Z) = sup\{min\{\rho_Y(u), \rho_Z(v)\}: u \geq v\} \qquad (15)$$

for any pair $(\tilde{R}_Y, \tilde{R}_Z)$ of fuzzy expected return rates.

In the next step we determine multicriteria comparison $\mathcal{W}\epsilon \mathbb{Y} \times \mathbb{Y}$ by maximization fuzzy expected return rates and by the minimization variance. Formed in this way relation we describe as the predicate $\check{Y} \supseteq \check{Z}$ which reads

$$\text{\textit{Security }} \check{Y} \text{ \textit{is no more effective than security }} \check{Z}. \qquad (16)$$

In a formal way this multicriteria comparison is defined by equivalence

$$\check{Y} \supseteq \check{Z} \Leftrightarrow \tilde{R}_Y \succcurlyeq \tilde{R}_Z \wedge \sigma_Y \leq \sigma_Z. \qquad (17)$$

In this situation the relation $\mathcal{W}$ is fuzzy preorder defined by its membership function $\nu_W: \mathbb{Y} \times \mathbb{Y} \to [0,1]$. For any pair of securities $\check{Y}, \check{Z} \epsilon \mathbb{Y}$ mentioned membership function is represented by the identity

$$\nu_W(\check{Y}, \check{Z}) = \begin{cases} \nu_Q(\tilde{R}_Y, \tilde{R}_Z) & \sigma_Y \leq \sigma_Z \\ 0 & \sigma_Y > \sigma_Z \end{cases}. \qquad (18)$$

The set $\tilde{\Phi}$ of effective securities is equal to the Pareto optimum defined by multicriteria comparison (17). The set $\tilde{\Phi}$ is represented by its membership function $\varphi: \mathbb{Y} \to [0,1]$ determined by the identity

$$\varphi(\check{Y}) = inf\{max\{\nu_W(\check{Y}, \check{Z}), 1 - \nu_W(\check{Z}, \check{Y})\}: \check{Z} \epsilon \mathbb{Y}\}. \qquad (19)$$

The value $\varphi(\check{Y})$ is interpreter as a truth value of the sentence:

$$\text{\textit{The security }} \check{Y} \text{ \textit{is effective}}. \qquad (20)$$

In this way behavioural reasons for investment decision making were applied for description similarity individual securities to effective one. This result was obtained without the assumption that the probability distribution of return rates is gaussian. Presented here normative theory explains that the divergence between the normative investor's purpose and the real goal of investment strategy is implied by behavioural aspects of financial market perception. Each explained paradox is apparent one. This formal theory allows to control of the choice of securities similar to the effective one. It follows from the fact, that using this theory we can to determine the truth value of the sentence (20).

We described above the case, when the investor determines the effective securities, taking into account only the risk of uncertainty. Now we focus our attention on the case, when the investor simultaneously takes into account the uncertainty risk and the imprecision risk.

Let us consider now the multicriteria comparison $\mathcal{L}\epsilon\mathbb{Y}\times\mathbb{Y}$ determined by maximization fuzzy expected return rates and three criteria of minimization described above risk measures. Formed in this way relation we describe as the predicate $\check{Y} \sqsupseteq \check{Z}$ which reads

$$\textit{Security } \check{Y} \textit{ is no more strictly effective than security } \check{Z}. \tag{21}$$

In a formal way this multicriteria comparison is defined by equivalence

$$\check{Y} \sqsupseteq \check{Z} \Leftrightarrow \widetilde{R}_Y \succcurlyeq \widetilde{R}_Z \wedge \sigma_Y \leq \sigma_Z \wedge \delta_Y \leq \delta_Z \wedge \varepsilon_Y \leq \varepsilon_Z. \tag{22}$$

In this situation the relation $\mathcal{L}$ is fuzzy preorder defined by its membership function $\nu_L: \mathbb{Y} \times \mathbb{Y} \to [0,1]$. For any pair of financial instruments $\check{Y}, \check{Z}\epsilon\mathbb{Y}$ mentioned membership function is represented by the identity

$$\nu_L(\check{Y},\check{Z}) = \begin{cases} \nu_Q(\widetilde{R}_Y, \widetilde{R}_Z) & \sigma_Y \leq \sigma_Z \wedge \delta_Y \leq \delta_Z \wedge \varepsilon_Y \leq \varepsilon_Z \\ 0 & \sim (\sigma_Y \leq \sigma_Z \wedge \delta_Y \leq \delta_Z \wedge \varepsilon_Y \leq \varepsilon_Z) \end{cases}. \tag{23}$$

The set $\widetilde{\Psi}$ of strictly effective securities is determined as the Pareto optimum defined by multicriteria comparison (22). The set $\widetilde{\Psi}$ is represented by its membership function $\psi: \mathbb{Y} \to [0,1]$ determined by the identity

$$\psi(\tilde{Y}) = inf\{\max\{\nu_L(\tilde{Y},\tilde{Z}), 1 - \nu_L(\tilde{Z},\tilde{Y})\}: \tilde{Z}\epsilon\mathbb{Y}\}. \tag{24}$$

The value $\psi(\tilde{Y})$ is interpreted as a truth value of the sentence:

$$\textit{The security } \check{Y} \textit{ is strictly effective.} \tag{25}$$

Investing only in strictly efficient securities can be recognized as normative investor's goal. This strategy causes rejection of such investment alternatives which are admittedly attractive from the viewpoint of classical Markowitz theory, but unfortunately gathered information about them is imprecise.

If investors considers purchase or sale the security $\check{Y}$ then they can take into account the values $\varphi(\tilde{Y})$ and $\psi(\tilde{Y})$. Investors should limit the area of their investments to securities characterized by relatively high value of these indicators. Also investors should limit the sale of their securities to those for which mentioned above indicators have low values. Presented by Piasecki (2011) considerations suggest that in same time individual investors use different values of these indicators. This diversification follows from diversification of subjective, behavioural reasons for investment decisions.

## 5. Final remarks

Applications of presented above normative model are involving several difficulties. The main difficulty is the high formal and computational complexity of the tasks of determining the membership function of effective securities set. Computational complexity of the normative model is the price which we pay for the lack of detailed assumptions about the return rate. On the other side, low logical complexity is important good point of presented in this paper formal model.

The problem of finding a membership function of effective securities set also can be based on econometric analysis of financial markets.

## References

Czogala E., Gottwald S. and Pedrycz W., (1982), Contribution to application of energy measure of fuzzy sets, Fuzzy Sets and System 8, pp 205-214

Gottwald S., Czogała E. and Pedrycz W., (1982), Measures of fuzziness and operations with fuzzy sets, Stochastica 6, pp 187-205

Hiroto K., Concepts of probabilistic sets, Fuzzy Sets and Systems 5, 1981, pp 31-46

Kahneman D., Tversky A., Prospect theory: an analysis of decision under risk, Econometrica 47, 1979, pp 263-292.

Markowitz H.S.M. , Portfolio selection, Journal of Finance 7,1952, pp 77-91.

Piasecki K., Fuzzy P-measures and their application and decision making, [in]: Kacprzyk J., Fedrizzi M. [eds.] Combining Fuzzy Imprecision with Probabilistic Uncertainty, Lecture Notes in Economics and Mathematical Systems 310, Springer Verlag, Berlin, 1988, pp 39-54

Piasecki K., Behavioural present value, Behavioural & Experimental Finance eJournal 4/2011, 2011